\documentclass[prl,floatfix]{revtex4} 
\usepackage{graphicx}
\usepackage{float}
\usepackage{psfrag}

\begin{document}
\def\t{temperature }
\def\tn{temperature}
\def\dm{density matrix }
\def\dmn{density matrix}
\def\ady{adiabaticity }
\def\adyn{adiabaticity}
\def\ad{adiabatic }
\def\adn{adiabatic}
\def\de{decoherence }
\def\den{decoherence}
\def\sy{system }
\def\syn{system}
\def\wv{wavefunction }
\def\wvn{wavefunction}
\def\hn{Hamiltonian}
\def\h{Hamiltonian }
\def\px{$\phi^{ext}$}
\def\sy{system }
\def\syn{system}
\title{%
\vskip 36pt Simulations of ``Decoherence'' with Noise Pulses}
\thanks{Contribution to the special issue of {\it Molecular
Physics} in honor of R. A. Harris.}

\author{L.~Stodolsky}
\affiliation{Max-Planck-Institut f\"ur Physik 
(Werner-Heisenberg-Institut),
F\"ohringer Ring 6, 80805 M\"unchen, Germany}

\date{Nov. 2005}

\begin{abstract} 
A simulation of \de as random noise in the \h is studied.
The full \h for the rf Squid is used, with the parameters chosen
such that there is a double-potential well 
configuration where the two quasi-degenerate lowest levels are well
separated from the rest. The results for these first two levels
are in quantitative agreement with expectations from the  ``spin
1/2'' picture for the behavior of a two-state \syn.
\end{abstract}

\maketitle

 Back around 1980 Bob Harris and I wrote the first
papers~\cite{us},~\cite{us1}   
in the field which in the meantime has come to be called ``\de''.
My title should really be `` Some Simulations of `Quantum Damping'
which is what
 Bobby and I called it in those days, but you can't fight MGM. 
 Briefly, the message was--and is--that
issues often thought to belong to  the arcane domain of
``measurement theory'' could actually lead to effects of physical
significance and could be  calculated concretely.

We were trying to understanding  why experiments we had proposed
for the L$\leftrightarrow$R oscillations of chiral
molecules~\cite{osc}
-- still  beautiful experiments waiting to be done--- weren't as
easy as they sounded. The enormous sensitivities theoretically
possible  were obviously too good to be true, but we wanted to
understand the reason. The answer turned out  to be  ``\de''.

The \dm for a two-state \sy like (L,R) is a Hermitian  2x2  matrix,
and since such a matrix can always be
 written in terms of
the Pauli matrices $\sigma$ we introduce three parameters $\bf P$ 
 and write 
\begin{equation}\label{rhomata}
\rho= 1/2 (1+{\bf P}\cdot {\bf \sigma}) \; .
\end{equation}
 $\bf P$ may be looked at as a ``polarization vector'' in an
abstract ``spin space'' and contains
the information
on the coherence of the two states. This language allows 
a helpful visualization as a ``spin 1/2 \sy ''--where the spin or
polarization point of course in the abstract  space--, and for the
time development we study the motions of $P$.
  For this we have a ``Bloch-like'' equation

\begin{equation}\label{pdota}
\dot {\bf P}={\bf P}\times {\bf V}- D\,{\bf P}_{T}\;.
\end{equation}
 V represents the internal \h and the damping parameter D 
the effects of the environment. D can change the length of P, while
V cannot, and
so for example can make a pure state ($\vert  {\bf P}\vert=1 $)
into a
mixed state ($\vert  {\bf P}\vert<1 $). Furthermore, as study of
this equation shows, a
large D  inhibits the natural V-induced rotations of $P$, and so
can  stop or seriously slow down things like
L$\leftrightarrow$R oscillations. Thus in addition to clarifying
why the experiments
aren't easy, ``quantum damping'' could also explain the permanence
of optical isomers
~\cite{aa}.

[ The usual notation is such that the abstract ``z-axis''
corresponds to
the property in question, so that $1/2(1+P_z)$ is the probability 
of finding L and $1/2(1-P_z)$ the  probability  of finding R. ${\bf
P}_{T}$, where
``T'' stands for ``transverse''
means the components of ${\bf P}$ perpendicular to the ``z-
direction'', that is the x,y
components; it
represents the degree of phase coherence between the two basis
states. In Eq~\ref{pdota} we have   taken the random external
perturbations   to be along the   abstract z-axis, causing
stochastic rotations around that axis. This corresponds to a low
temperature situation where there is no direct barrier hopping
between the two states. ]

 Given Eq~\ref{pdota}, the important question becomes the
calculation of D. We found  a formula for it, resembling a kind
of off-diagonal optical theorem, in terms of the S-matrices for the
environmental atoms or molecules scattering on our
\sy\cite{us},~\cite{us1}. However
Bobby likes to understand things in more than one way and in the
appendix to ref~\cite{us} he gave a little model in terms of
random pulses rotating a spin.

Over the years this formalism has been generalized and applied in
various fields~\cite{rev}. Recently we have been working on the
presently hot
topic of mesoscopic devices and quantum computing, where for the
latter
subject \de is the major if not overriding issue. In particular we
have gone into some
detail for the rf Squid, showing for example how to perform the
logic operations NOT and CNOT~\cite{sq}. Thus we have been carrying
out
numerical simulations for these Squid-based \syn s, and  recently
we've begun to try  simulating the effects of \de for the devices.
This simulation is in some ways like the picture with random pulses
in  our old paper--a slightly fancier version with the pulses in
the \hn, and I thought it might be amusing to present it
here.

The rf Squid   is described by a \h where the ``position
coordinate''  is $\phi$, the flux  
in the Squid.  In certain parameter ranges the potential in the \h
 has  a
double-potential well form, as is shown in Fig 1.

 The \h is:

\begin{equation}\label{hama1}
H={-1\over 2\mu }{\partial ^2\over\partial \phi^2}
 +V_0\{{1\over 2}[(\phi-\phi^{ext})^2 ]+\beta\, cos\phi\}\;. 
\end{equation}
Fig 1 also shows  the first four energy
levels.
The parameters $\mu$, $\beta$ and $V_0$ are related to the
properties of the Squid. The quantity \px ~is an external flux 
we can apply and vary to carry out our various operations, 
essentially it is used to raise and lower the relative heights of
 the two potential wells.
We can adjust the parameters such that the two lowest states of
this \hn, at \px =0 split only by the small tunneling energy,
effectively
constitute a two-state \sy relatively well isolated from the
other states of the \hn . Thus  it should be describable by the
formalism of Eqns~\ref{rhomata},\ref{pdota}, if we look only at the
two lowest states. 

\begin{figure}
{{\includegraphics[width=0.35\hsize,angle=-90]{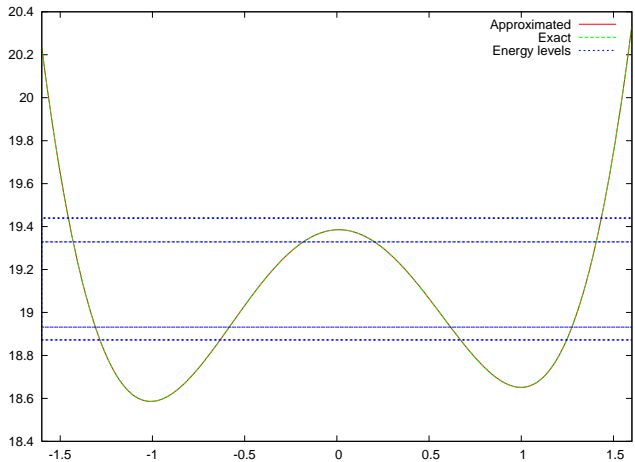}}}
\caption{An example of the potential in Eq~\ref{hama1}  with
\px=0.0020. The first four energy levels from a numerical solution
of the stationary Schroedinger equation are indicated. Varying \px
makes the potential asymmetric and allows manipulation of the
levels and quantum gate operations.}
\end{figure}

We introduce \de into the \sy by supposing a random noise in the \h
and then evolving an initial \wv to a final \wv $\psi^a$ with the
\h $H^a$ with a given realization of the noise $a$. We then 
obtain the \dm as an average over \wvn s from different
realizations

\begin{equation} \label{av}
\rho=\overline{\psi \psi^{\dagger}}={1\over N}\sum_{a=1}^N {\psi^a
\psi^{a\dagger}}\; .
\end{equation}

The noise $\cal N $ is introduced into the \h as a kind of flux
noise--which may be physically the most relevant-- by sending

\begin{equation}\label{noi}
\phi^{ext} \to \phi^{ext}+{\cal N}^a(t)\;,
\end{equation}
so that for each realization of the noise ${\cal N}^a(t)$ we have
some time dependent \h  $H^a$. This \h is then used to evolve the
\wv to obtain $\psi^a$.

In the computer the noise is generating by random pulses of
magnitude
$ \pm \Delta$. The  frequency of the noise is then governed by the
effective time between sign switches.
Analysis~\cite{gat} of the effect of this in the \h leads to the
conclusion
that for the effective two-state \sy composed of the two lowest
states we should have 

\begin{equation}\label{d21}
 D=4 (V_0 \phi_c)^2 \int_0^{\infty}\overline{{\cal N}(t){\cal
N}(0)} dt=4 (V_0 \phi_c)^2 {\Delta^2\over \omega_c}\; ,
\end{equation}
where $\omega_c $ is the frequency of the noise and $ \phi_c$ is
the ``coordinate position'' where the \wv tends to be localized,
usually  $\phi_c\approx 1$, as can be seen from  Fig 1.
 \begin{figure}
{{\includegraphics[width=0.5\hsize]
{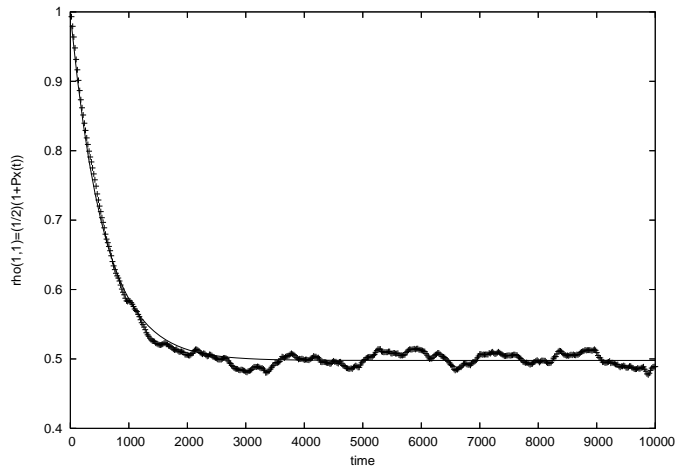}}}
\caption{Effect of  decoherence as simulated with random  noise in
the  \hn, with \px =0.   A
pure state at t=0 is converted to the maximally mixed state for the
two-state \syn. The
exponential decay is $e^{-Dt}$ is  fit (thin line )with
$D=0.00175$, while the prediction from Eq~\ref{d21} is D=0.00164.
The quantity plotted is the probability of finding the original
state, which is  $\rho_{11}$ in the basis of energy
eigenstates or $(1/2)(1+P_x)$. Noise parameters
were $\Delta=0.00032,\omega_c=.05$. }
\end{figure}
The numerical evolution of the time dependent Schroedinger equation
is carried out with a fast algorithm  using algebraic manipulations
in an harmonic oscillator basis~\cite{wos}.  
 The \dm resulting from the average over \wvn s $\psi^a({\phi})$ is
then given in the 
``position coordinates'' $\sim \rho(\phi',\phi)$. This is, however,
not immediately in the form Eq~\ref{rhomata}. Although we hope that
if we start in one of the two lowest states we stay there, we are
now dealing with a many-state \syn, in principle
 containing components from all the states of the Squid \h
Eq~\ref{hama1}.
Therefore, depending on the question being asked,  the results must
be evaluated 
by finding the matrix elements of $ \rho(\phi',\phi)$ in
some basis of \wvn s.

 A first simple question we can ask is if we get the right behavior
and the right  D. For this purpose we can start with an energy
eigenfunction of  the symmetric \px=0 \h and examine the
probability to find this state at a later time. The only
energy splitting is due to the small tunneling (only $V_x\neq 0$)
and the starting situation corresponds to $\bf P$ along the x-axis.
Turning on the  noise, we get Fig 2 for the evolution of the \dm
element
for the probability of finding the initial state, that is
 $\rho_{1,1}$ in the  basis of the original energy eigenstates,
$\rho_{1,1}=(1/2)(1+P_x)$.
\begin{figure}[h]
\centerline{{\includegraphics[width=0.5\hsize]
{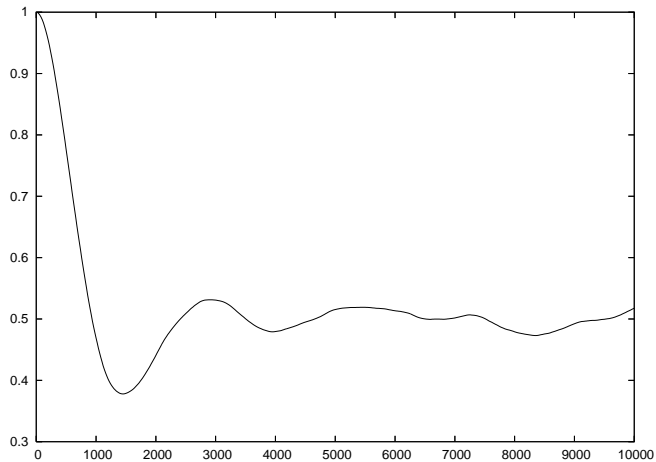}}}
\caption{Same conditions as Fig 2 but with the initial state in the
``up'' or ``z-
direction''. Thus the vertical axis corresponds to $(1/2)(1+P_z)$.}
\end{figure}
 Gratifyingly, we see that $\rho_{11}$ drops off exponentially to
the value of
1/2, as Eq~\ref{pdota} would predict. Furthermore the value of D
from the fit
shown in the plot is D=0.00175, while the prediction from
Eq~\ref{d21} with
the noise parameters used would be D= 0.00164. Also, although
excitations of the higher states of the Squid were possible, the
relaxation to 1/2 and not less shows that such excitations were
small, at least for the parameters used here (they can be produced
by using  large $\Delta$). This and the good agreement with
Eq~\ref{d21} appears to show that, at least in this parameter
range, the effective spin 1/2 picture using the first two states
works well.

We can now try something a little more sophisticated with the same
\hn, and start with an initial state where $P$ points ``up'', that
is in the abstract ``z-direction''. Since in this symmetric
configuration the energy eigenstates ``point'' in the x-direction,
this is no longer an energy eigenstate
and in the  absence of \de we  would expect  $P$ to simply rotate
in the z-y plane, around the x-axis. Fig 3 shows what happens, with
the same parameters as for Fig 2. As would be expected, the
oscillations and the damping combine to give damped oscillations. 
We hope to present more details and applications in the near
future~\cite{gat}.

In conclusion I should perhaps stress that simulations such as
these are 
only phenomenological and are not meant to replace calculations
with
 the real basic physical  amplitudes that determine D. 
For example, the phase~\cite{aa}, \cite{ph} arising in the full S-
matrix treatment
does not show up here. 
But of
course the simulations can be quite useful in understanding what is
 to be expected for our  logic gates. In any event our
 quantum damping is alive and well,
and continues to find interesting and ever-widening applications.

\vskip0.75cm
 The numerical work is based on the methods of J. Wosiek~\cite{wos}
as
developed into a very useful program package by A. G\"orlich  and
 P. Korcyl, students at the Jagellonian University, Cracow.


\end{document}